\documentclass[conference]{IEEEtran}
\IEEEoverridecommandlockouts

\usepackage{cite}
\usepackage{amsmath,amssymb,amsfonts}
\usepackage{algorithmic}
\usepackage{algorithm}
\usepackage{graphicx}
\usepackage{textcomp}
\usepackage{xcolor}
\def\BibTeX{{\rm B\kern-.05em{\sc i\kern-.025em b}\kern-.08em
    T\kern-.1667em\lower.7ex\hbox{E}\kern-.125emX}}
\begin{document}

\title{Adaptive Entanglement Management in Quantum Multi-Core Architectures  \\
}

\author{
    \IEEEauthorblockN{
        Rajeswari Suance P S\IEEEauthorrefmark{1},
        Anubhab Dutta\IEEEauthorrefmark{1}
        Ruchika Gupta\IEEEauthorrefmark{2},
        John Jose\IEEEauthorrefmark{1}
    }
    \IEEEauthorblockA{
        \IEEEauthorrefmark{1}Indian Institute of Technology Guwahati, India \quad
        \IEEEauthorrefmark{2}Chandigarh University, India \quad
    }
    \IEEEauthorblockA{
        \textnormal{\{s.rajeshwari, d.anubhab, johnjose\}@iitg.ac.in, ruchikae7396@cumail.in}
    }
}


\maketitle

\begin{abstract}

Scalable quantum computing architectures increasingly rely on multi-core designs, where qubits are distributed across multiple processing cores interconnected through a quantum Network-on-Chip (NoC). In such systems, inter-core communication is typically realized through entanglement-assisted quantum teleportation, making efficient entanglement generation critical for performance. In this paper, we perform a comparative study of three entanglement management paradigms for multi-core quantum processors: reactive on-demand generation (ODG), proactive continuous pre-generation (CGP), and an adaptive continuous pre-generation approach (ACGP). While ODG generates entanglement only when required, CGP reduces average teleportation latency by pre-generating EPR pairs in the background. To improve upon this, we propose ACGP which dynamically adjusts entanglement generation probabilities based on observed inter-core communication patterns. We evaluate these approaches using an extended SeQUeNCe simulator on mesh-based multi-core architectures on real benchmark circuits. Results show that ACGP significantly reduces average teleportation latency compared to ODG and CGP. Although pre-generation introduces fidelity degradation due to storage time, entanglement purification effectively restores fidelity with minimal impact on latency. These results demonstrate that adaptive entanglement managements can substantially improve communication efficiency in scalable quantum multi-core systems.

\end{abstract}


\section{Introduction}
Quantum computing is evolving from the current monolithic Noisy Intermediate-Scale Quantum (NISQ) processors to distributed multi-core architectures \cite{smith_micro22, alarcorn_iscas23}. NISQ devices house tens to hundreds of qubits on a single chip. Quantum computers offer significant potential for applications in complex molecular simulations, optimization problems, and cryptography. However, to fully harness their power, the number of qubits in a system must scale \cite{preskill2018quantum}. Increasing the number of qubits on a monolithic chip leads to severe crosstalk, requires complex wiring, and reduces manufacturing yield \cite{escofet2023interconnect}. To address these challenges, multi-core quantum computing architectures have been proposed, where each core contains tens to hundreds of qubits, and these cores are interconnected via a quantum-classical Network-on-Chip (NoC). Scalability is achieved by increasing the number of cores, thus mitigating issues of monolithic architectures.

Quantum algorithms are represented by quantum circuits consisting of quantum gates. The input qubits for multi-qubit gates must be adjacent for the gate to execute. If the interacting qubits reside on different cores, inter-core communication becomes necessary \cite{rodrigo2021modelling}. This is a significant deviation from classical systems, as qubits are inherently fragile. They can decohere if exposed to the environment, making direct transfer of qubits impractical. Instead, quantum teleportation is used, which leverages quantum entanglement and classical communication to transfer qubits without physically moving them. The most time-consuming aspect of teleportation is the generation of entangled pairs at the source and destination \cite{cacciapuoti2020entanglement}. This process is probabilistic and often requires multiple attempts to create a single entanglement, depending on the physical hardware parameters.

The most intuitive approach for generating entanglement is to initiate it when an inter-core teleportation request is received, known as on-demand entanglement generation \cite{suance2025decentralized}. While this approach exists, it introduces high latency. To address this, we propose the pre-generation of entanglement across different cores, allowing these entangled pairs to serve inter-core communication requests immediately when needed. Our major contributions are as follows:
\begin{enumerate}
    \item We propose a random Continuous Pre-Generation protocol (CGP) for pre-generating entanglements, to be used in future inter-core communications.
    \item We propose an enhanced version of CGP, which adapts to communication patterns and learns from them, called the Adaptive Continuous Pre-Generation Protocol (ACGP).
    \item We make a detailed comparative study of on-demand generation (ODG) with CGP and ACGP in terms of average teleportation latency and entanglement fidelity. We test these protocols on real benchmarks and demonstrate the improvements and trade-offs between them.
\end{enumerate}
The remainder of this paper is organized as follows: Section II provides background on quantum computing, multi-core architectures, and quantum teleportation, motivating the entanglement generation bottleneck. Section III discusses the baseline on-demand generation protocol along with the proposed continuous and adaptive continuous generation protocols. Section IV presents the experimental setup and results, followed by a discussion of the findings. Finally, Section V concludes the paper and outlines future research directions.

\section{Background}
\subsection{Quantum Computing Fundamentals}
A qubit is the unit of quantum information. Unlike a classical bit, which have definite value of 0 or 1, a qubit can exist in superposition of both states. This property enables quantum parallelism and forms the basis of quantum computational advantage. Qubits are highly susceptible to environmental interactions that induce decoherence and degrade quantum information. Consequently, most physical implementations operate at cryogenic temperatures to preserve coherence.

Another fundamental property of qubits is entanglement. In an entangled system, the quantum state of one qubit is correlated with that of another, regardless of their physical separation. Measurement collapses a qubit’s state to its classical bit. An arbitrary quantum state cannot be copied, a property of quantum mechanics formalized by the no-cloning theorem. These characteristics fundamentally distinguish quantum computation from classical information processing.

Quantum algorithms are represented as quantum circuits composed of single-qubit and two-qubit gates. Gates that operate on disjoint sets of qubits are executed in parallel and are grouped into layers. While operations within a layer occur concurrently, there exists a strict sequential dependency between layers. Furthermore, two-qubit gates require the participating qubits to be physically adjacent on the hardware. Limited connectivity necessitates additional routing operations to satisfy architectural constraints. The transformation of a logical circuit into a hardware-compliant circuit through qubit movement and gate insertion is referred to as quantum circuit mapping \cite{russo2025telesabre}. In multi-core systems, such movement operations may induce inter-core communication overhead.

\subsection{Multi-Core Quantum Architectures}

Quantum algorithms are offloaded from a host computer operating at room temperature to quantum processing units maintained at cryogenic temperature \cite{escofet2023interconnect}. Current NISQ devices follows a monolithic architecture. However, scalability limitations in qubit count, control wiring, and fabrication yield motivate the transition toward modular and multi-core architectures. The multi-core quantum architecture partitions qubits across multiple interconnected cores. Each core comprises computational qubits dedicated to algorithm execution and communication qubits reserved for inter-core state transfer. To support coordinated operation, these systems employ a hybrid  NoC design that handles both classical and quantum information \cite{suance2025decentralized}. The classical NoC transmits classical bits required to generate control signals, synchronize quantum operations, and support inter-core qubit transfer. The quantum NoC transfers photons emitted from communication qubits to establish entanglement. This entanglement serves as fundamental resource for inter-core quantum communication \cite{rodrigo2021modelling}.

\subsection{Inter-Core Communication: Quantum Teleportation}

As already mentioned, direct physical transfer of qubits between cores is impractical due to decoherence and hardware constraints. To enable interactions between qubits located in different cores, several techniques are employed, including the insertion of SWAP gates, qubit shuttling in trapped-ion systems, and quantum teleportation using entanglement property \cite{palesi2024assessing}. Among these approaches, inter-core communication in multi-core quantum architectures commonly relies on entanglement-assisted protocols.
\begin{figure}
    \centering
    \includegraphics[width=0.6\linewidth]{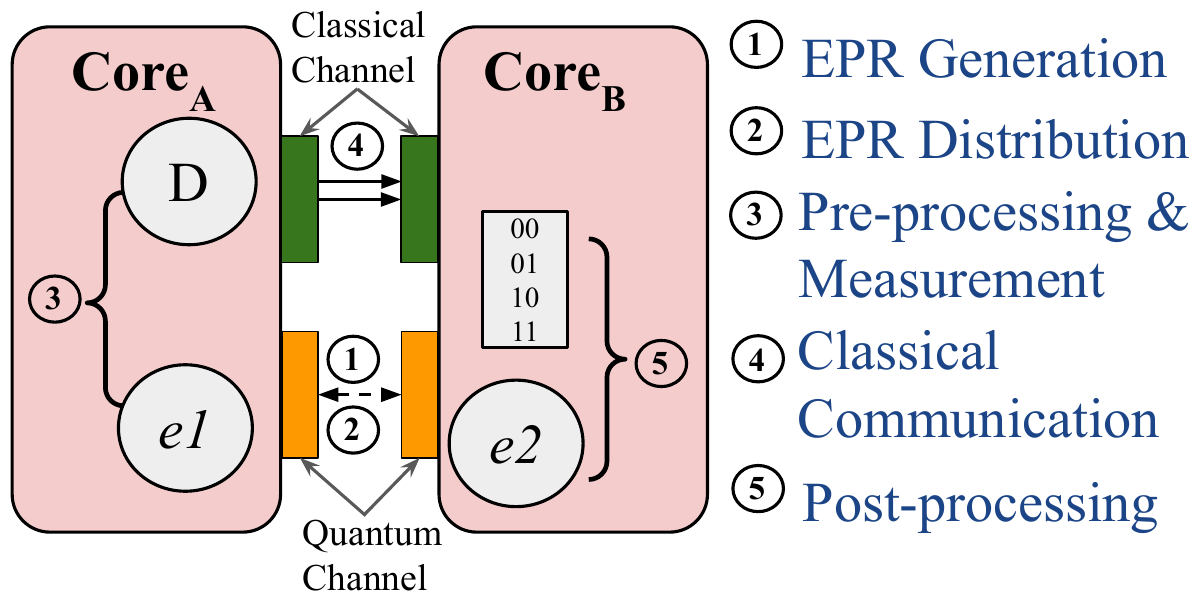}
    \caption{Quantum Teleportation with steps}
    \label{fig:teleportation}
\end{figure}

Quantum teleportation depends on quantum entanglement and classical communication and requires an entangled pair shared between the source and destination cores. This process is illustrated in Figure~\ref{fig:teleportation}. First, entanglement pair also called as EPR pairs are generated and distributed to both cores (Steps 1 and 2). At the source, pre-processing operations and measurements are performed on the data qubit $D$ and one qubit of the EPR pair $e1$, collapsing the state into classical bits, as illustrated in step 3. These bits are transmitted to the destination core through the classical NoC, as shown in step 4. At the destination, the quantum state is reconstructed by applying appropriate post-processing operations to the second EPR qubit $e2$ and the received classical bits, as depicted in Step 5. The original state is destroyed during measurement, ensuring compliance with the no-cloning theorem.

Two variants of quantum teleportation are commonly used for inter-core communication. When the input qubits of a two-qubit gate are located in different cores, either of two approaches can be adopted \cite{russo2025telesabre}. In the first approach, known as teledata, the quantum state of one qubit is teleported to the destination core, after which the two-qubit gate resumes execution locally. In the second approach, remote gate operations are performed using entanglement and classical communication without transferring the quantum state. In this case, the initial mapping between logical and physical qubits remains unchanged. Gates are then executed within the respective cores with the help of entanglement.

The performance of these approaches is largely determined by the availability of entangled pairs. As reported in prior work \cite{rodrigo2021modelling}, the mean EPR pair generation time is $10^3 ns$, which is far higher than EPR distribution ($0.01 ns$), pre-processing ($390 ns$), post-processing ($30 ns$), and classical communication ($0.02 ns$), highlighting that the generation of entangled pairs is both necessary and the key enabler for faster inter-core quantum teleportation.

\subsection{Entanglement Generation Bottleneck}

Entanglement can be generated in three ways \cite{cacciapuoti2020entanglement}. The first method generates an EPR pair at a central node between two cores. The second method generates the EPR pair at the source, which then travels through a quantum channel to the destination cores. The third method generates the EPR pair at both endpoints.

Based on these  approaches, multi-core architectures can be broadly classified into two categories. The first category employs a centralized EPR generator \cite{rodrigo2021modelling}. In this setup, each core is connected to the EPR generator via light-to-matter interfaces. However, the point-to-point connections and the centralized generator create a bottleneck, limiting the simultaneous generation and distribution of entanglement.
\begin{figure}
    \centering
    \includegraphics[width=0.6\linewidth]{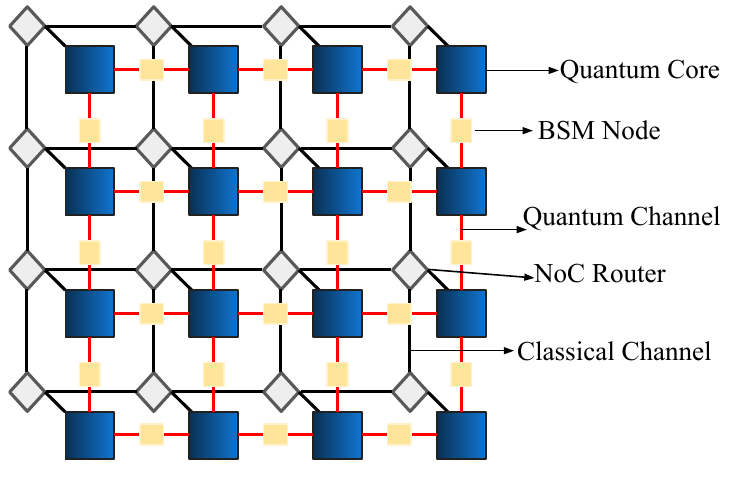}
    \caption{\centering Multi-Core Quantum System with Distributed Entanglement Generation \cite{suance2025decentralized}}

    \label{fig:multicore}
\end{figure}
The second category is a decentralized design, as shown in Figure~\ref{fig:multicore} \cite{suance2025decentralized}. Here, entanglement is generated at both endpoints. Entanglement is established with the help of an intermediate Bell State Measurement (BSM) node positioned between cores, typically using the Barrett–Kok entanglement generation protocol. We use this model in the paper for all studies. 

To support communication between non-neighboring cores, entanglement swapping extends elementary entanglement links to longer distances. Furthermore, when the fidelity of entangled pairs degrades due to noise and decoherence, entanglement purification is applied to improve their quality. In this process, two or more low-fidelity entangled pairs are combined to produce a single entangled pair with higher fidelity \cite{jnane2022multicore}.

\section{Methodology}

Quantum teleportation requires one EPR pair for each inter-core qubit transfer. Since entanglement generation is inherently probabilistic, multiple attempts may be required before a high-fidelity EPR pair is successfully established. For complex quantum algorithms that require frequent inter-core communication, a continuous supply of high-fidelity entanglement becomes essential. Delays in entanglement generation can stall teleportation, increasing overall algorithm runtime \cite{chakraborty2019distributed}. Such delays negatively affect computational qubits due to decoherence and may increase the overall error probability. Therefore, the policy used for entanglement generation plays a critical role in system performance. Based on how and when entanglement is generated, we categorize entanglement management into three paradigms.

\subsection{On-Demand Entanglement Generation (ODG)}

ODG is a reactive paradigm where entanglement is generated only when an inter-core communication request arrives \cite{abane2025entanglement}. Upon receiving a request, deterministic routing (e.g., XY routing) reserves communication resources along the path using the classical NoC. Adjacent cores then begin probabilistic entanglement generation. Once neighboring EPR pairs are successfully established, intermediate cores perform entanglement swapping to construct a long-distance EPR pair between the source and destination cores. Teleportation proceeds only after this process completes.

While straightforward to implement, this approach increases communication latency because entanglement generation occurs sequentially and probabilistically. Under high traffic conditions, processors may stall while waiting for EPR generation to succeed. These delays increase total execution time and exacerbate decoherence in computational qubits.
\begin{algorithm}[t]
\caption{Protocol at Initiator}
\label{alg:init-node-bg}
\begin{algorithmic}[1]
\WHILE{true}
    \STATE Sleep for $T_{\text{rand}}$
    \IF{$used\_memory < memory\_limit$}
        \STATE Reserve memory $m_A$
        \STATE $used\_memory \gets used\_memory + 1$
        \STATE Select neighbor N according to probability distribution P
        \STATE Send \textsc{QUERY} to $N$
        \STATE Wait for reply
        \IF{reply = \textsc{YES}}
            \STATE Trigger \textsc{Entanglement\_Generation}$(m_A)$
        \ELSE
            \STATE Release memory $m_A$
            \STATE $used\_memory \gets used\_memory - 1$
        \ENDIF
    \ENDIF
\ENDWHILE
\end{algorithmic}
\end{algorithm}

\begin{algorithm}[t]
\caption{Protocol at Neighbor}
\label{alg:nb-node-bg}
\begin{algorithmic}[1]
\WHILE{true}
    \STATE Wait for \textsc{QUERY} from another node
    \IF{$used\_memory < memory\_limit$}
        \STATE Reserve memory $m_B$
        \STATE $used\_memory \gets used\_memory + 1$
        \STATE Send \textsc{YES} reply to Node-A
        \STATE Start \textsc{Entanglement\_Generation}$(m_B)$
    \ELSE
        \STATE Send \textsc{NO} reply to Node-A
    \ENDIF
\ENDWHILE
\end{algorithmic}
\end{algorithm}

\subsection{Continuous Pre-Generation (CGP)}

CGP is a proactive paradigm for reducing entanglement generation latency, where cores pre-generate EPR pairs in the background using idle communication qubits. In a naive implementation, a core with $k$ neighbors selects a neighboring core with probability $P = \frac{1}{k}$ and attempts entanglement generation. Successfully generated EPR pairs are stored locally. When an inter-core communication request arrives, the local resource manager checks whether a pre-generated EPR pair exists. If available, analogous to a cache hit in classical systems, generation is skipped and swapping followed by teleportation occurs immediately, effectively eliminating generation latency.

To enable this behavior, a background protocol continuously performs entanglement pre-generation at each node. Each node imposes a limit on the fraction of communication qubits that can be allocated to this protocol. The protocol, shown in Algorithms~\ref{alg:init-node-bg} and~\ref{alg:nb-node-bg}, operates using two finite-state machines that manage entanglement requests. The initiator node periodically wakes up, checks memory availability and allocation limits, and, if resources permit, reserves a memory unit and sends a request to a neighboring node. The neighbor evaluates its own resources and either accepts the request initiating entanglement generation or rejects it. Upon receiving the response, the initiator either proceeds with entanglement generation or releases the reserved memory and returns to sleep.

Although pre-generation significantly reduce latency when a matching EPR pair is available, it introduces inefficiencies. Because this strategy does not consider actual spatio-temporal communication patterns, EPR pairs may be generated between cores that rarely or never communicate. Given the limited capacity of quantum memory, storing unused entangled pairs reduces effective resource utilization. Furthermore, entangled pairs degrade over time due to decoherence. Continuous monitoring of fidelity is therefore required, and low-fidelity pairs must be discarded, leading to additional regeneration overhead. Consequently, random pre-generation may waste resources and fail to align with actual communication demand.

\subsection{Adaptive Continuous Pre-Generation (ACGP)}

ACGP is an intelligent paradigm that addresses the inefficiencies of random pre-generation. It uses an adaptive, learning-based mechanism for EPR pre-generation.  This paradigm extends CGP by dynamically adjusting neighbor selection based on observed communication behavior. Instead of uniformly selecting neighboring cores, each core maintains a probability distribution that reflects the likelihood of communicating with each neighbor. These probabilities are updated using historical inter-core communication patterns. By learning routing statistics and request frequency, the system prioritizes neighbors with higher expected traffic. This adaptive policy increases the likelihood that pre-generated EPR pairs will be utilized in future teleportation operations. As a result, it improves the probability of latency-free communication while reducing wasted entanglement resources and unnecessary regeneration.

In our implementation, each core maintains a table that associates neighboring cores with communication probabilities. This table is periodically updated based on recent traffic patterns generated by the circuit. Each node records incoming communication requests, and whenever a new pre-generation event occurs, the table is updated to reflect the recent history of entanglement usage. During an update event, the protocol examines recent entanglement paths within a time window. For each neighbor $i$, the probability is updated as

\begin{equation}
p_i' =
\begin{cases}
p_i + \Delta, & i \in U \\
p_i - \Delta, & i \in V \\
p_i, & \text{otherwise}
\end{cases}
\end{equation}

where $U$ denotes the set of neighbors appearing in recent entanglement paths, $V$ denotes neighbors with unused pre-generated entanglement, and $\Delta$ is a fixed update step. The updated probabilities are then normalized as

\begin{equation}
p_i^{new} = \frac{p_i'}{\sum_{j \in N} p_j'}
\end{equation}

where $N$ is the set of neighboring cores.

\begin{figure*}[t]
    \centerline{\includegraphics[width=\textwidth]{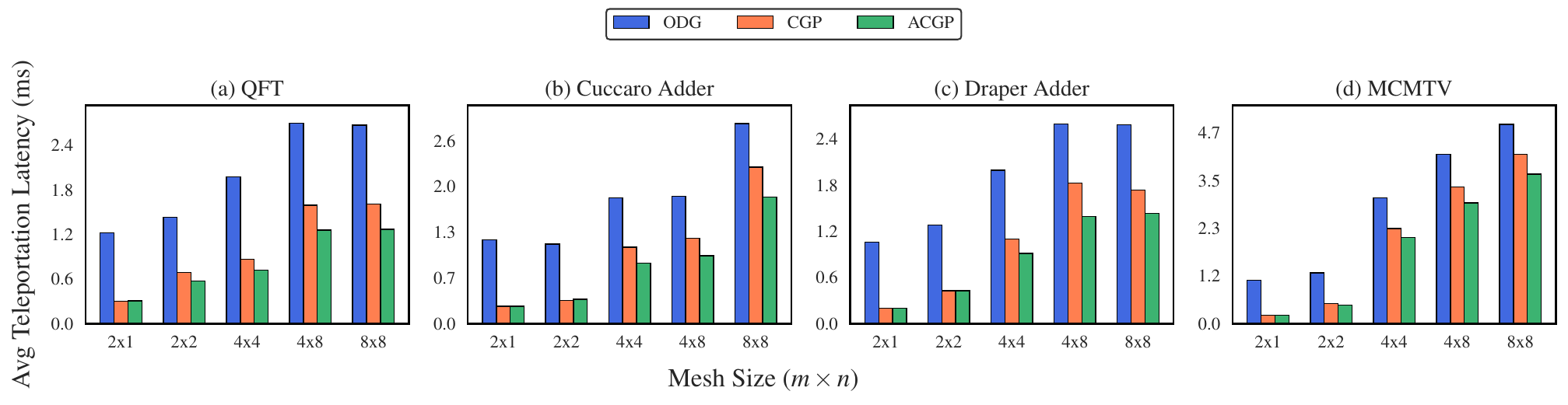}}
    \caption{Comparison of average teleportation latency for proposed protocols across different mesh topologies}
    \label{fig:latency}
\end{figure*}

\begin{figure*}[t]
\centerline{\includegraphics[width=\textwidth]{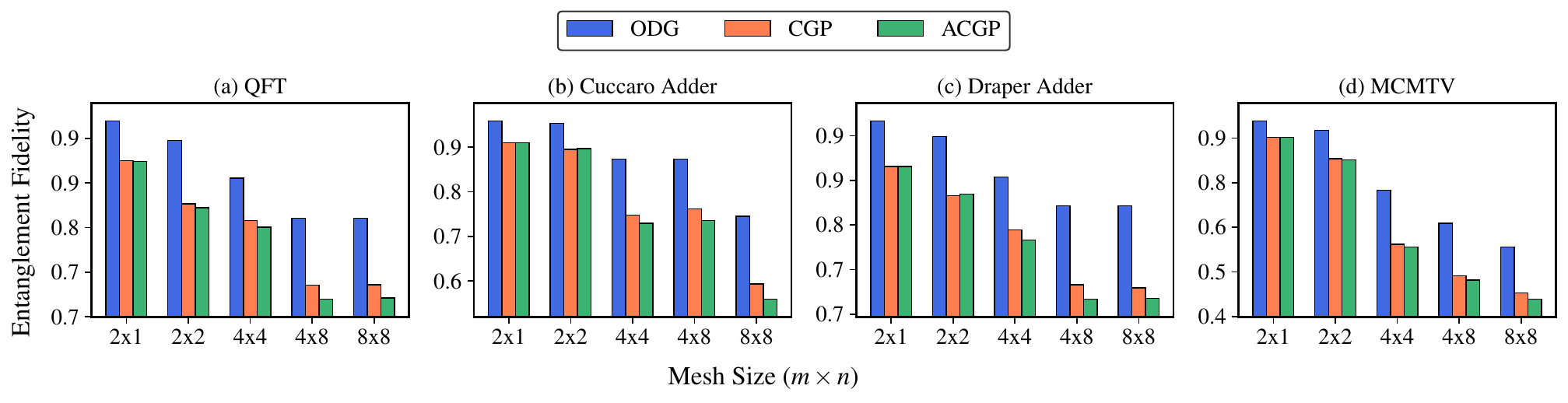}}
\caption{Comparison of entanglement fidelity for proposed protocols across different mesh topologies}
\label{fig:fidelity}
\end{figure*}

\begin{table}[t]
\caption{System Parameters}
\label{tab}
\begin{center}
\begin{tabular}{|c|c|}
\hline
\textbf{Parameter} & \textbf{Value} \\
\hline
Computation Qubits per core & 8 \\
\hline
Communication Qubits per core & 4 \\
\hline
Swapping Probability & 0.95 \\
\hline
Coherence Time & 1 ms \\
\hline
Memory Efficiency & 0.85 \\
\hline
BSM Detection Efficiency & 0.95 \\

\hline
\end{tabular}
\end{center}
\end{table}

\section{Experiments}

We evaluate our approach using the multi-core quantum architecture proposed \cite{suance2025decentralized}. We extend the SeQUeNCe quantum network simulator \cite{wu2021sequence} to support a multi-core architecture connected through a mesh NoC. A mesh topology is selected because it provides a practical balance between routing complexity and physical fabrication feasibility. Inter-core communication is implemented as an entanglement-mediated teleportation process. Each core contains 8 computation qubits and 4 communication qubits. For an $m \times n$ mesh topology, the total number of computational qubits is therefore $m \times n \times 8$. We conduct the experiments across a range of mesh topologies, starting with smaller configurations like $2 \times 1$ and $2 \times 2$, and extending to larger meshes such as $4 \times 8$ and $8 \times 8$.

To model realistic hardware constraints, we use fixed physical parameters summarized in Table~\ref{tab}. Memory efficiency ($\eta_{mem}$) represents the probability that a communication qubit successfully absorbs and retrieves a photon during entanglement generation. The BSM detector efficiency ($\eta_{BSM}$) represents the probability that detectors correctly register incoming photons at BSM nodes. The swap success probability represents the likelihood that intermediate cores successfully extend entanglement through entanglement swapping. Finally, the coherence time of communication qubits defines the lifetime of stored entangled pairs, after which unused pairs are discarded to maintain fidelity.

To evaluate the proposed entanglement pre-generation protocols, we map representative quantum subroutines including the Quantum Fourier Transform (QFT) \cite{coppersmith2002approximate}, Cuccaro Adder \cite{cuccaro2004new}, Draper Adder \cite{draper2000addition}, and MCTMV onto the mesh topology. These circuits are generated using \texttt{qiskit} \cite{contributors2023qiskit}. We assume all-to-all qubit connectivity within each core, which simplifies intra-core communication and allows us to focus on inter-core communication. Logical qubits are mapped directly to physical qubits, where each logical qubit is assigned to a specific core denoted as $\lfloor C_i \rfloor$, with $i$ representing the core identifier.

An inter-core communication request occurs when the input qubits of a two-qubit gate reside in different cores. All inter-core communications within a circuit layer are executed in parallel, while dependencies between layers enforce sequential execution. This behavior creates bursty communication traffic, where some layers generate many teleportation requests while others generate few. We evaluate system performance using two primary metrics: average teleportation latency and entanglement fidelity.  

\subsection{Impact on Average Teleportation Latency}

Average teleportation latency is defined as the time between the initiation and completion of an inter-core teleportation request. Figure~\ref{fig:latency} illustrates the comparison of average teleportation latency for the proposed protocols. Compared to ODG, the execution of QFT across different topologies results in an average latency reduction of 53\% for CGP and 61\% for ACGP. Similarly, the execution of the Cuccaro adder shows a reduction of 49\% for CGP and 57\% for ACGP, relative to ODG. The execution of Draper adder and MCTMV also exhibits comparable trends. These results demonstrate that ODG incurs higher latency because entanglement generation begins only after a teleportation request is issued. When the circuit transitions from layers with few teleportation requests to layers with many requests, the bursty communication pattern leads to increased waiting time for entanglement generation.

ACGP achieves the lowest average teleportation latency because it pre-generates entanglement opportunistically during idle computation periods and during layers with fewer communication requests. By learning historical communication patterns, ACGP prioritizes neighbors that are more likely to require entanglement, increasing the probability that an EPR pair is already available when needed. Although CGP also reduces average teleportation latency compared to ODG, it performs less efficiently than ACGP because its neighbor selection is uniform and does not adapt to varying communication patterns.

\subsection{Impact on Entanglement Fidelity}
The latency benefits of entanglement pre-generation come at the cost of reduced entanglement fidelity. Entanglement fidelity measures how closely the generated entangled state approximates an ideal Bell state after accounting for noise and decoherence. A higher fidelity indicates better quality entanglement. Figure~\ref{fig:fidelity} presents a comparison of the fidelity of consumed entanglement pairs for the proposed protocols across various mesh topologies. ODG achieves the highest fidelity because entanglement is used immediately after generation, minimizing exposure to decoherence. In contrast, CGP and ACGP store pre-generated entangled pairs that may remain idle before being used. During this waiting period, decoherence reduces fidelity. Across benchmarks, ACGP exhibits an average fidelity degradation of approximately 10\% relative to ODG, except for MCTMV, where the degradation reaches 16\%. Larger mesh topologies, further amplify this effect because longer routing distances increase waiting time before entanglement consumption.

\subsection{Restoring Fidelity with Purification}
To mitigate fidelity degradation while preserving latency improvements, we apply entanglement purification to the CGP and ACGP protocols. Purification combines multiple lower-fidelity entangled pairs to generate a single higher-fidelity pair.

\begin{figure}[t]
\centering
\includegraphics[width=0.75\linewidth]{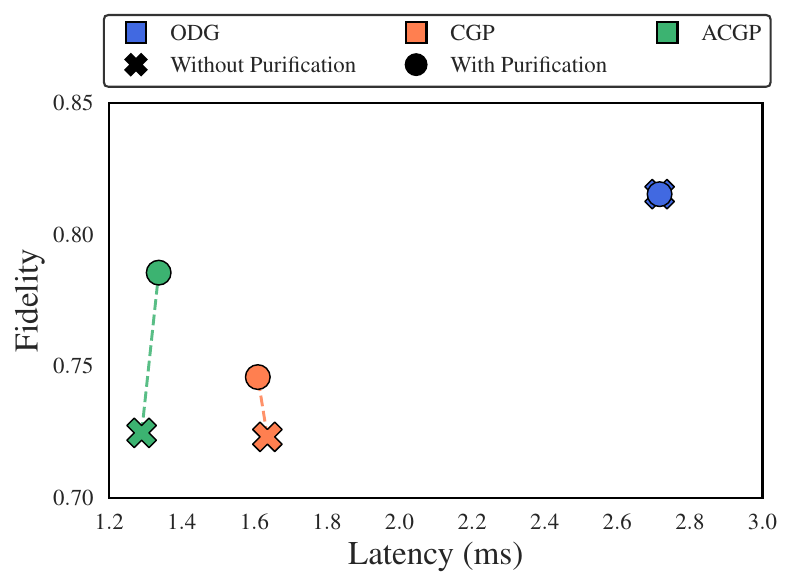}
\caption{Latency–fidelity trade-off achieved through purification}
\label{tradeoff}
\end{figure}

Figure~\ref{tradeoff} demonstrates that purification enhances fidelity without significantly affecting latency in the scalable $8 \times 8$ mesh topology. Since purification runs in the background, its overhead is minimal, as the required classical communication and Bell-state measurements occur simultaneously with computation. For the execution of Cuccaro and Draper adders, the fidelity degradation decreases from approximately 12\% to about 6\% relative to ODG after applying purification. Notably, since ODG already operates with high-fidelity pairs, purification does not result in further improvements for this protocol.

\section{Conclusion}

In this paper we presented entanglement management strategies for multi-core quantum processors interconnected through a mesh NoC. We compared reactive ODG, proactive CGP, and an adaptive ACGP. While CGP reduced latency through background entanglement generation, its random neighbor selection resulted in inefficient entanglement utilization. ACGP addressed this limitation by adapting entanglement generation probabilities based on observed communication patterns between cores. Experimental results across several quantum benchmarks showed that ACGP achieved the lowest average teleportation latency while maintaining acceptable fidelity. Although pre-generation introduced fidelity degradation due to storage time, purification effectively restored fidelity with minimal latency overhead.
 
These results indicated that adaptive entanglement managements significantly improved communication efficiency in scalable quantum multi-core systems. Future work can include exploring multi-hop virtual entanglement links and adaptive routing strategies that exploit partially available entangled connections to further improve latency and resource utilization.

\section*{Acknowledgment}
The authors gratefully acknowledge the funding support from ISEA Project Phase III, Ministry of Electronics and IT, Govt. of India.

\bibliographystyle{IEEEtran}
\bibliography{reference}

@article{preskill2018quantum,
  title={{Quantum Computing in the NISQ era and beyond}},
  author={Preskill, John},
  journal={Quantum},
  volume={2},
  pages={79},
  year={2018},
  publisher={Verein zur F{\"o}rderung des Open Access Publizierens in den Quantenwissenschaften}
}

@inproceedings{suance2025decentralized,
  title={{Decentralized Framework for Teleportation in Quantum Core Interconnects}},
  author={Suance, PS Rajeswari and Gupta, Ruchika and Palesi, Maurizio and Jose, John},
  booktitle={IEEE Computer Society Annual Symposium on VLSI (ISVLSI)},
  pages={1--6},
  year={2025},
}

@inproceedings{escofet2023interconnect,
  title={{Interconnect Fabrics for Multi-Core Quantum Processors: A Context Analysis}},
  author={Escofet, Pau and Rached, Sahar Ben and Rodrigo, Santiago and Almudever, Carmen G and Alarc{\'o}n, Eduard and Abadal, Sergi},
  booktitle={IEEE/ACM 16th International Workshop on Network on Chip Architectures (NoCArc)},
  pages={34--39},
  year={2023}
}

@inproceedings{palesi2024assessing,
  title={{Assessing the Role of Communication in Scalable Multi-Core Quantum Architectures}},
  author={Palesi, Maurizio and Russo, Enrico and Patti, Davide and Ascia, Giuseppe and Catania, Vincenzo},
  booktitle={IEEE International Symposium on Embedded Multicore/Many-core Systems-on-Chip (MCSoC)},
  pages={482--489},
  year={2024}
}

@inproceedings{russo2025telesabre,
  title={{TeleSABRE: Heuristic Layout Synthesis in Multi-Core Quantum Systems with Teleport Interconnect}},
  author={Russo, Enrico and Vinciguerra, Elio and Palesi, Maurizio and Patti, Davide and Ascia, Giuseppe and Catania, Vincenzo},
  booktitle={IEEE International Conference on Quantum Computing and Engineering (QCE)},
  pages={749--758},
  year={2025},
}

@INPROCEEDINGS{alarcorn_iscas23,
  author={Alarcón, Eduard and Abadal, Sergi and Sebastiano, Fabio and Babaie, Masoud and Charbon, Edoardo and Bolívar, Peter Haring and Palesi, Maurizio and Blokhina, Elena and Leipold, Dirk and Staszewski, Bogdan and Garcia-Sáez, Artur and Almudever, Carmen G.},
  booktitle={IEEE International Symposium on Circuits and Systems (ISCAS)}, 
  title={Scalable multi-chip quantum architectures enabled by cryogenic hybrid wireless/quantum-coherent network-in-package}, 
  year={2023},
  pages={1--5},
}

@INPROCEEDINGS{smith_micro22,
  author={Smith, Kaitlin N. and Ravi, Gokul Subramanian and Baker, Jonathan M. and Chong, Frederic T.},
  booktitle={55th IEEE/ACM International Symposium on Microarchitecture (MICRO)}, 
  title={{Scaling Superconducting Quantum Computers with Chiplet Architectures}}, 
  pages={1092--1109},
  year={2022},
}

@inproceedings{rodrigo2021modelling,
  title={{Modelling Short-range Quantum Teleportation for Scalable Multi-Core Quantum Computing Architectures}},
  author={Rodrigo, Santiago and Abadal, Sergi and Almud{\'e}ver, Carmen G and Alarc{\'o}n, Eduard},
  booktitle={Eight Annual ACM International Conference on Nanoscale Computing and Communication (NANOCOM)},
  pages={1--7},
  year={2021}
}

@article{cacciapuoti2020entanglement,
  title={{When Entanglement Meets Classical Communications: Quantum Teleportation for the Quantum Internet}},
  author={Cacciapuoti, Angela Sara and Caleffi, Marcello and Van Meter, Rodney and Hanzo, Lajos},
  journal={IEEE Transactions on Communications},
  volume={68},
  number={6},
  pages={3808--3833},
  year={2020},
}

@article{chakraborty2019distributed,
  title={{Distributed Routing in a Quantum Internet}},
  author={Chakraborty, Kaushik and Rozpedek, Filip and Dahlberg, Axel and Wehner, Stephanie},
  journal={arXiv preprint arXiv:1907.11630},
  year={2019}
}

@article{wu2021sequence,
  title={{SeQUeNCe: A Customizable Discrete-Event Simulator of Quantum Networks}},
  author={Wu, Xiaoliang and Kolar, Alexander and Chung, Joaquin and Jin, Dong and Zhong, Tian and Kettimuthu, Rajkumar and Suchara, Martin},
  journal={Quantum Science \& Technology},
  volume={6},
  number={4},
  pages={045027},
  year={2021},
  publisher={IOP Publishing}
}

@article{coppersmith2002approximate,
  title={{An approximate Fourier Transform useful in Quantum Factoring}},
  author={Coppersmith, Don},
  journal={arXiv preprint quant-ph/0201067},
  year={2002}
}

@article{cuccaro2004new,
  title={{A New Quantum Ripple-carry Addition Circuit}},
  author={Cuccaro, Steven A and Draper, Thomas G and Kutin, Samuel A and Moulton, David Petrie},
  journal={arXiv preprint quant-ph/0410184},
  year={2004}
}

@article{draper2000addition,
  title={{Addition on a Quantum Computer}},
  author={Draper, Thomas G},
  journal={arXiv preprint quant-ph/0008033},
  year={2000}
}

@article{contributors2023qiskit,
  title={{Qiskit: An Open-source Framework for Quantum Computing}},
  author={Contributors, Qiskit},
  journal={Zenodo: Geneva, Switzerland},
  year={2023}
}

@article{jnane2022multicore,
  title={{Multicore Quantum Computing}},
  author={Jnane, Hamza and Undseth, Brennan and Cai, Zhenyu and Benjamin, Simon C and Koczor, B{\'a}lint},
  journal={Physical review applied},
  volume={18},
  number={4},
  year={2022},
  publisher={APS}
}

@article{abane2025entanglement,
  title={{Entanglement Routing in Quantum Networks: A Comprehensive Survey}},
  author={Abane, Amar and Cubeddu, Michael and Mai, Van Sy and Battou, Abdella},
  journal={IEEE Transactions on Quantum Engineering},
  year={2025},
}

\end{document}